\documentclass[rnote]{aa} % for the research notes
\usepackage{graphicx}
\usepackage{color}
\usepackage{txfonts}
\usepackage{url}
\newcommand{\xmm}{{\em XMM-Newton}}
\newcommand{\cxo}{{\em Chandra}}

\newcommand{\xicma}{\mbox{$\xi^1$\,CMa}}
\newcommand{\bcep}{\mbox{$\beta$\,Cep}}
\newcommand{\bcru}{\mbox{$\beta$\,Cru}}

\def\changed{}
\begin{document}

   \title{On X-ray pulsations in $\beta$ Cephei-type variables \thanks{The 
scientific results reported in this article are based
on observations made by the {\em Chandra} and {\em XMM-Newton} 
X-ray Observatories, data obtained 
from the {\em Chandra} and {\em XMM-Newton} Data Archives, 
and observations made by the {\em Chandra} and {\em XMM-Newton} 
and published previously in cited articles}}

   %\subtitle{Research Note}

   \author{L. M. Oskinova
          \inst{1}
          \and
          H. Todt\inst{1}
          \and
          D. P. Huenemoerder\inst{2}
          \and 
          S. Hubrig\inst{3}
          \and 
          R. Ignace\inst{4}
          \and
          W.-R. Hamann\inst{1}
          \and
          L. Balona\inst{5}}
\institute{Institute of Physics and Astronomy, University of Potsdam, 
              14476 Potsdam\\
              \email{lida@astro.physik.uni-potsdam.de}
         \and
         Massachusetts Institute of Technology, Kavli Institute 
     for Astrophysics and Space Research, 70 Vassar St., Cambridge, MA 02139, USA
          \and  
    Leibniz Institute for Astrophysics Potsdam (AIP), An der 
Sternwarte 16, 14482 Potsdam, Germany
          \and        
    Department of Physics and Astronomy, East Tennessee State 
    University, Johnson City, TN 37663, USA
           \and 
    South African Astronomical Observatory, PO Box 9, 
    Observatory 7935, Cape Town, South Africa }
   \date{Received September 15, 1996; accepted March 16, 1997}

% \abstract{}{}{}{}{} 
% 5 {} token are mandatory
 
  \abstract
  {\bcep-type variables are early B-type stars that are characterized by 
   oscillations observable in their optical light curves. 
   At least one  \bcep-variable also shows periodic variability in X-rays. 
   } 
   {Here we study the X-ray light curves in a sample  of
    \bcep-variables to  
    investigate how common X-ray pulsations are for this type of stars.}
   {We searched the {\em Chandra} and {\em XMM-Newton} X-ray archives 
    and selected stars that were observed by these telescopes for at 
    least three  optical pulsational periods. 
    We retrieved and analyzed the X-ray data for \object{$\kappa$\,Sco}, 
    \object{$\beta$\,Cru}, and \object{$\alpha$\,Vir}. The X-ray light curves 
    of these objects were studied to test for their variability and periodicity.}
   {While there is a weak indication for X-ray variability in \object{$\beta$\,Cru}, 
   we find no statistically significant evidence of X-ray pulsations in any 
   of our sample stars. This might be due either to the insufficient data 
   quality or to the physical lack of modulations.  New, more sensitive observations 
   should settle this question.}
   {}


   \maketitle
%
%________________________________________________________________

\section{Introduction}

Oscillating stars can be found almost everywhere in the HR  diagram. The
hottest and most massive  oscillating stars are
$\beta$\,Cephei-type variables.  Born with masses between $8\,M_\odot$
and  $18\,M_\odot$, and while still young and burning hydrogen in their
cores, these B0--B2  type stars pulsate with periods of a few hours. 
The physical mechanism  that drives these oscillations is understood well
and attributed to changes in the opacity inside 
the star during the pulsation cycle  \citep[``$\kappa$-mechanism'',][]{dz1993}.

Like other hot massive stars, \bcep-type variables drive stellar  winds
by their intense UV radiation. Photons that are scattered  or absorbed
in spectral lines transfer their momentum and thus accelerate  the
matter to supersonic velocities.  This driving mechanism is unstable. It
is generally  believed that  the wind instability results in shocks 
where  part of the wind material  is  heated to  X-ray emitting
temperatures  \citep[e.g.,][]{feldmeiera1997}.  

In $\beta$\,Cep-type variables, the deposition of mechanical energy 
from stellar pulsations  can provide additional heating of the inner
wind  regions. This idea was put forward to explain the observed excess
in the extreme  UV spectrum of  \object{$\beta$\,CMa}  \citep{cas1996}.  
Recent work on classical  Cepheids \citep{Neilson2008, Engle2014} has
demonstrated that pulsations may  power X-ray  emission even from
such cool stars.  Thus, it seems reasonable to assume that stellar 
oscillations are {\changed closely linked} with the X-ray production 
in \bcep\ variables.  

Recently, it has been found that X-rays from the strongly magnetic \bcep-variable 
\object{$\xi^1$\,CMa} pulsate in phase  with the optical 
light curve{\footnote{\url{http://sci.esa.int/xmm-newton/54101/}}
\citep{osk2014}. To investigate whether  \xicma\ is a unique  
anomaly or if other \bcep-variables are also X-ray pulsators, we  searched the 
X-ray archives for observations of \bcep-variables. The archival data for a 
selected sample of variable  were analyzed to extract the X-ray light curves
(see Figure\,1) and study their variability.   
In this {\it Research Note} we present the results of this study. 

%=============================================================
\begin{figure}[t!]
\centering
\includegraphics[width=0.99\columnwidth]{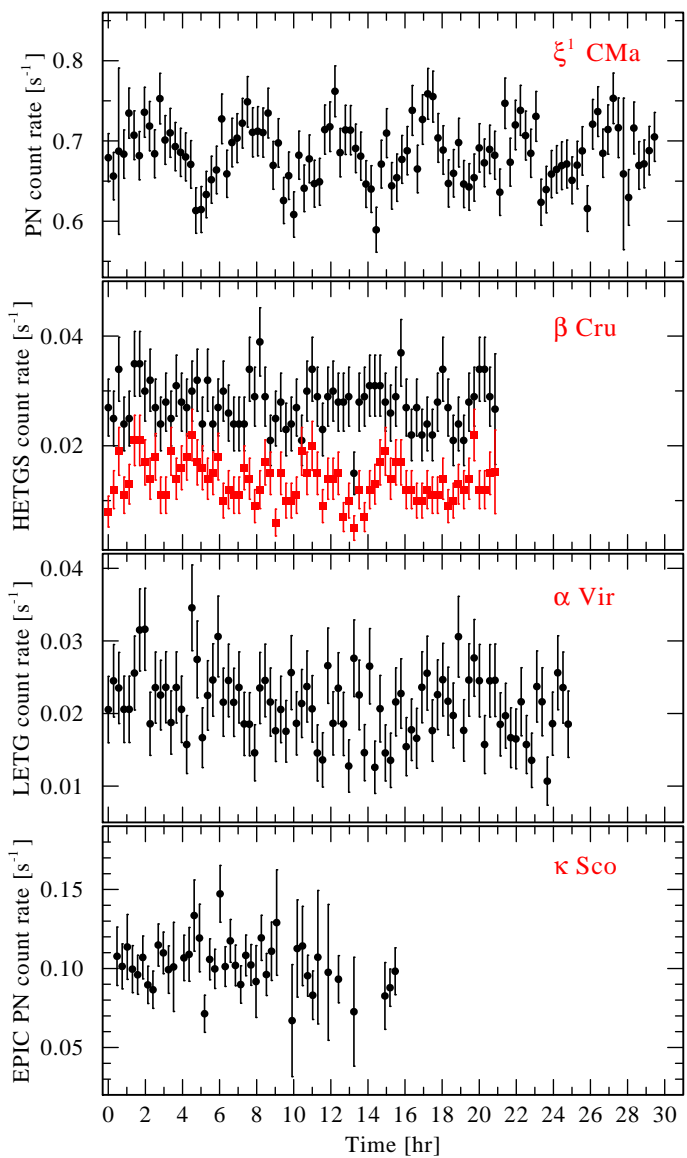}
\caption{X-ray light curves in  0.2--10.0\,keV 
band of \bcep-type variables binned by 1000\,s. The count rate of \xicma\ is at least an order of magnitude 
higher than for the other stars.   
The horizontal axis shows the time after the start of the observation. 
Vertical bars represent 1$\sigma$ errors. From top to bottom:
the EPIC PN light curve of \xicma; the HETG (added +1 and -1 order) 
light curve of \object{$\beta$\,Cru}, where the lower red light curve is in the hard 
band (1.0-10.0\,keV) where the detection of pulsations was claimed 
by \citet{coh2008}; LETG (added +1 and -1 order) light curve 
of \object{$\alpha$\,Vir}; EPIC PN light curve of \object{$\kappa$ Sco}.}
\label{fig:lc}
\end{figure}
%-------------------------------------------------------------

\section{Previous investigations of X-ray pulsations in \bcep-variables}

Despite being a numerous and astrophysically important class of objects, 
\bcep-variables have not been studied well in X-rays.  The first X-ray survey
of six $\beta$\,Cep-type stars was performed with  the {\em Einstein} 
observatory \citep{agr1984}. No correlation was found between X-ray  and 
pulsational, rotational, or binary properties. 

The interest in X-ray properties of \bcep-variables was renewed with the 
launch of the {\em Rosat} X-ray observatory. Four \bcep\ stars were  
among the sample of 12 nearby near-main-sequence B-stars observed 
by  {\em Rosat} \citep{cas1994}. Timing ana\-lysis was performed for 
all targets, but no variability was found except those attributed 
to the spacecraft wobble.   

%%%%%%%%%%%%%%%%%%%%%%%%%%%%%%%%%%%%%%%%%%%%%%%%%%%%%%%%%%%
%++++++++++++++++++++++ Table 2 ++++++++++++++++++++
\begin{table*}
%\centering
\caption{\xmm\ and {\em Chandra} observations of 
 \bcep-variables covering at least three stellar pulsational periods}
%\vspace{2mm}
\label{tab:res}
\begin{tabular}{lccccccccc}
\hline \hline
  & Sp & Period$^{\rm A}$ & Obs & Counts$^{\rm B}$ &  $n_{\rm p}^{\rm C}$  & X-ray period? 
 & $A_{\rm max}^{\rm D}$  &  Comment$^{\rm E}$ & Reference  \\ 
    &    & [h] & & $P_{\rm puls}^{-1}$ & & & & & \\  \hline 
\object{$\beta$ Cen}  & B1III    & 3.684 & XMM & 25000& 6  & no  & $< 3$\% & mag &
\citet{raas2005} \\ 
\object{$\beta$ Cep}  & B2IIIe   & 4.572 & XMM & 2500 & 9  & no  & $< 6$\% & mag &
\citet{fav2009} \\ 
\object{$\beta$ Cru}  & B0.5IV   & 4.589 & CXO & 225$^{\rm F}$ & 4  & no & $< 26^{\rm F}$\% 
& non-mag & this work \\
\object{$\kappa$ Sco} & B1.5III  & 4.795 & XMM & 1700  & 3  & no  & $< 11$\% & ? &
this work \\
\object{$\alpha$ Vir} & B1.5IV-V & 6.521 & CXO & 470   & 3  & no  & $< 21$\% & ? &
this work \\
\object{$\xi^1$ CMa}   & B0.5-1IV & 5.030 & XMM & 12800 & 6  & yes & $10$\% & mag &
\citet{osk2014} \\ \hline 
\hline  
\end{tabular} \\
A -- main period of stellar oscillations as listed in the 
``Catalog of Galactic $\beta$ Cephei stars'' \citep{st2005};\\
B -- number of counts per optical pulsation period (EPIC PN for \xmm);\\  
C -- number of pulsation periods covered by X-ray light curve; \\
D -- amplitude of X-ray pulsations or its upper limit for non-detections; \\
E -- magnetic field presence;\\
F -- in the hard band where the pulsations were suggested by \citet{coh2008};
%\vspace{-3mm}
\end{table*}
%\vspace{-1cm}
%%%%%%%%%%%%%%%%%%%%%%%%%%%%%%%%%%%%%%%%%%%%%%%%%%%%%%%%%%%

{\em Rosat} observations of \bcep-variables were further ana\-lyzed for 
the presence of X-ray pulsations by \citet{coh2000}, who {\changed mentions} 
detecting periodic X-ray variability in four 
$\beta$\,Cephei-type variables.  Specifically, for the star $\beta$\,Cephei, 
it was shown that the hypothesis of a constant source can be rejected 
with 98\%\  confidence \citep[see Fig.\,5  in][]{coh2000}. 
\citet{coh2000} reports that the X-ray  variability of $\beta$\,Cep is periodic 
with the same period as stellar pulsational period in the optical. 

The  binary system \bcep\ consists of a 
primary magnetic B2III-type star \citep{don2001} 
and a secondary B6-B8e type star \citep{schnerr2006}. 
The magnetosphere of the primary was studied 
in detail by \citet{don2001}. It was predicted that an equatorial 
disk-like structure would be present around this star. A signature of such a 
disk was found using interferometry by \citet{nar2011}. 

The occultations of X-ray emission sites by the disk should lead to observable 
X-ray variability on the rotational time scale.
To test this hypothesis, the more sensitive \xmm\ and \cxo\ observations of \bcep\ 
were obtained. However, only limited evidence of a modulation in the X-ray 
emission was found \citep{fav2009}. The variability was observed at a 5\%\ level
but in anti-phase with the model predictions. Furthermore, \citet{fav2009} 
performed a dedicated time analysis of the \xmm\ light curves but did not 
uncover any periodicity,  thus questioning the results of 
\citet{coh2000}. It was concluded that ``the fact that Cohen (2000) did 
not perform a period search,  but rather assumed the period and fitted 
the amplitude and phase,  makes it likely that the period 
reported by Cohen (2000) is spurious''  \citep{fav2009}.

An assumed period was also fitted to the X-ray light curve 
\bcep-variable \object{$\beta$ Cru} obtained with  the {\em Chandra} 
X-ray observatory  \citep{coh2008}. These authors applied 
a Kolmogorov-Smirnov test to the
unbinned  photon arrival times, as well as a $\chi^2$ fitting of a constant
source  model to the binned light curve and found  no evidence of X-ray
variability. However, when a grid of test periods was fitted to the
phased photon arrival times, it was found that for the X-ray light curve 
in the hard band ($h\nu > 1$\,keV), the rejection probabilities for some 
of the assumed periods were quite low. This was the basis for suggesting 
that the periodic variability is detected with a period $P = 4.588$\,h, 
which is the primary pulsation  mode period in $\beta$\,Cru. However, 
\citet{aerts2014} give the dominant frequency in  
\object{$\beta$ Cru} as 5.95867\,d$^{-1}$ (or 4.0277\,h). 
Polarimetric measurements for $\beta$\,Cru did not reveal
any magnetic field \citep{hub2006}.
We revisit the archival X-ray observations of \object{$\beta$ Cru} in 
Section \ref{sec:bcru}.

%=============================================================
\begin{figure*}[t!]
\centering
\includegraphics[width=1.75\columnwidth]{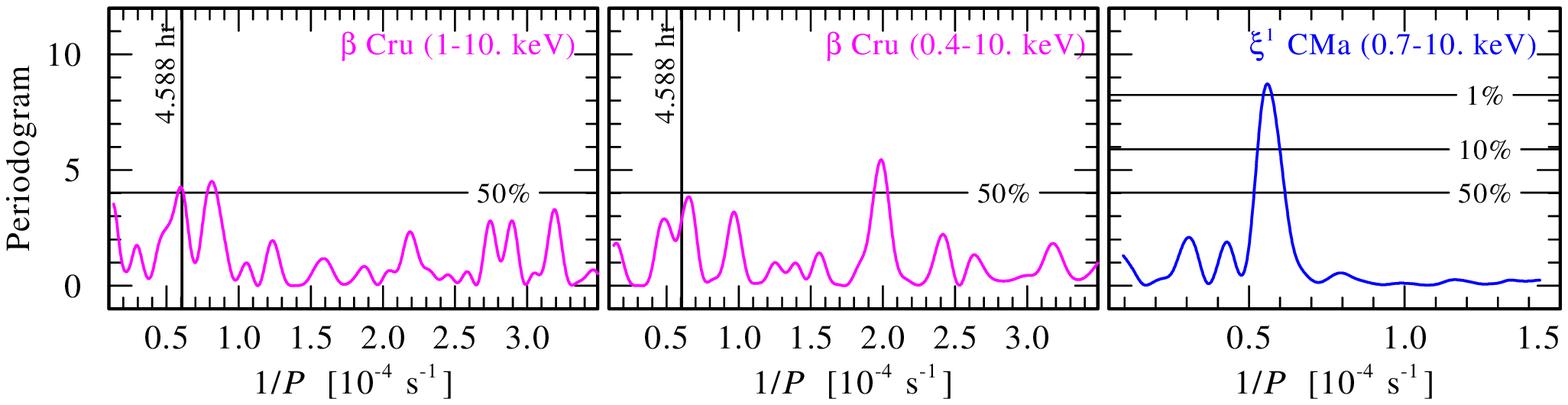}
\caption{Power density spectra based on the Fourier transform 
\citep{scargle1982, horne1986} of X-ray light curves.  
The periodogram was calculated from $\nu=1/P$ to 
$\nu=N_0/P$ with a spacing $P^{-1}$, where $N_0$ is the number 
of data points. Various false alarm probability levels are marked. 
From left to right: 1) combined zero and first-order
Chandra HETGS light curve of \bcru\ in hard (1.-10.0\,keV) 
band where pulsations were suggested by \citet{coh2008}. 
There is {\changed about a} 50\%\ probability that this period is spurious.
2) zero-order Chandra's HETGS light curve of \bcru\ in full 
(0.2-10.0\,keV) band, {\changed there is more than a 50\%\ probability 
that this period is spurious}; 3) EPIC PN light curve of \xicma\ in 
hard band (0.7-10.0\,keV).}
\label{fig:stat}
\end{figure*}
%-------------------------------------------------------------

Another \bcep-variable that has been studied in X-rays is \object{$\beta$ Cen}. This is
a binary system consisting of two $\beta$\,Cep-type 
stars of nearly equal mass 
with an orbital period of 357 days \citep[][and references therein]{aus2006}. 
The primary rotates significantly faster ($\varv\sin i =190\pm 20$\,km\,s$^{-1}$) 
than the secondary ($\varv\sin i =75\pm 15$\,km\,s$^{-1}$), as is evident from its 
rotationally broadened lines. The magnetic nature of the secondary, 
the slower rotating component in \object{$\beta$ Cen} is reported by \citet{alecian2011}. 

 Sensitive observations of \object{$\beta$\,Cen} with \xmm\ are reported 
by \citet{raas2005}. The X-ray light curves with different time binnings 
and energy bands were extracted from data obtained by all \xmm\ instruments 
and their combinations.  Scargle periodograms were produced but no 
periodicities with high statistical confidence found in any of the 
data sets. Moreover, no hint of periodicity with the period seen in optical
light curves was found even with low  statistical 
confidence. However, when the light curves were compared with the constant 
source hypothesis,  probabilities of 83\%\ (for EPIC PN) and 49\%\ 
(for EPIC MOS) were determined for the source not being constant.

\citet{osk2011} considered a small sample of \bcep-variables with well 
known pulsational behavior and existing X-ray observations and searched for
correlations between pulsational and X-ray  properties. 
After confirming previous results known from the {\em Einstein} and 
{\em Rosat} works, no clear relations could be found. This could be 
due to  the very limited sample.  

The sample of \bcep-variable with X-ray light curves of sufficient quality 
is  very small. It is observationally challenging to obtain X-ray 
light curves of high enough quality to establish periodic variability 
with high statistical significance. The stellar oscillations periods 
are typically four to six hours. To confidently detect pulsations in a light 
curve, the observations should be long enough to cover a number of 
periods. (In this study we adopt the minimum  of three periods.)
At the same time, the source should be X-ray bright in order to have  
enough counts per time bin  for a meaningful 
statistical analysis. 
In Table\,\ref{tab:res} we summarize the objects  that satisfy the above criteria.

\section{Looking for the variability in X-ray light curves of \object{$\beta$~Cru}, 
\object{$\kappa$~Sco}, 
and \object{$\alpha$~Vir}}   
\subsection{\object{$\beta$ Cru}}
\label{sec:bcru}

We retrieved the archival \cxo\ X-ray observations of \object{$\beta$\,Cru} 
(see Table\,\ref{tab:obs}) and analyzed them using the latest  calibration. 
The X-ray light curves were extracted in  different energy bands as defined 
in \citet{coh2008}: full (0.4-10.0\,keV), soft (0.4-1.0\,keV ),  and hard 
(1.0-10.0\,keV). The light curves were binned in the  1000\,s and 
3600\,s bins, and examples of light curves are shown in Fig.\,1.

{\changed 
For HETG dispersed photon data, we computed light curves using the
ACIS-Grating-Light-Curve ({\it aglc}) program (used by the Chandra
Grating-Data Archive and Catalog processing
\citep[TGCat,][]{hun2011}), in order to also produce a curve in the
hard band, in addition to the default broad-band curve available from
TGCat.  Zeroth-order curves (not shown) were created with the CIAO
\citep{CIAO:2006} program, {\it dmextract}.  The program {\it aglc} 
computes rates with the
explicit exposure records for each CCD, using the user-specified
combination of grating types, orders, and bandpasses.  For relatively
low count-rate systems, such as those analyzed here, with time binning
comparable to the dither period and with broad energy ranges spanning
all CCDs, chip-to-chip differences and chip gap effects are
negligible.  Furthermore, for the dispersed photons, there is no
concern with light leaks, as there could be for zeroth orders of
optically bright objects.}

Using the  $\chi^2$ test, the X-ray light curves were compared with those 
expected from the constant source hypothesis. As an example, for the 
zeroth-order light curve in full band 
and binned by 1000\,s, the  probability value is $P \approx 0.168$, 
which implies that the probability that the observed variability is 
solely a chance occurrence is less than $\approx 17$\%,\, which is statistically 
insignificant. 
Thus, the null-hypothesis assuming that the source is constant cannot 
be rejected. On the other hand, for the same light curve but binned into 3000\,s, 
$P \approx 0.0044$, which is statistically significant at the level 
$\alpha =0.01$. In the hard band, for the combined zeroth- and first-order 
light curve,  $P \approx 0.06$ is statistically insignificant 
with $\alpha = 0.05$. 

%%%%%%%%%%%%%%%%%%%%%%%%%%%%%%%%%%%%%%%%%%%%%%%%%%%%%%%%%%%
%++++++++++++++++++++++ Table 1 ++++++++++++++++++++
\begin{table*}
\centering
\caption{Log of analysed observations}
%\vspace{2mm}
\label{tab:obs}
\begin{tabular}{lccccc}
\hline \hline \\
Object & Telescope & ObsID & Start time & Duration & Count rate \\
       &           &       &            & [ks]     &  [s$^{-1}$] \\ \hline
\bcru  & CXO/HETG & 2575  & 2002-02-01 & 75 & 0.03 (-1+1 order) \\
$\kappa$\,Sco  & XMM  &0503500201 &2008-03-12 & 64  &  0.1 (pn)   \\
$\alpha$\,Vir   & CXO/LETG & 4509  & 2004-03-09 & 90 &   0.02 (-1+1 order) \\ \hline \hline
  
\end{tabular} 

\end{table*}
%\vspace{-1cm}
%%%%%%%%%%%%%%%%%%%%%%%%%%%%%%%%%%%%%%%%%%%%%%%%%%%%%%%%%%%

Thus, similar to the case of \object{$\beta$ Cen}, there are marginal indications 
of source variability. To investigate the presence of periodic 
modulations further, we constructed Scargle periodograms, with some examples shown in 
Fig.\,2. A peak with a false alarm probability of about 50\%\ is present at the 
frequency close to the stellar pulsational period. However, this is not the 
most prominent feature in the periodograms. For comparison, we show in 
Fig.\,2 
a periodogram of the \xicma\ in 0.7-10.0\,keV (hard) band, where the peak 
at the stellar pulsation frequency has a false alarm probability less than 1\%.   
  
Thus, from reanalysis of the \cxo\ data on \object{$\beta$ Cru,} we must conclude 
that in the statistical sense, it is likely that the period reported by 
\citet{coh2008} is spurious. Better quality observations are needed to 
establish whether the periodicity is indeed present in $\beta$\,Cru's 
X-ray light curve. 

\subsection{X-ray variability of \object{$\kappa$ Sco}}
The spectroscopic binary system, $\kappa$ Sco, has 
two B-type components orbiting with a period of 195\,d.
The primary (B1.5III)  is responsible for the pulsational 
variability \citep{harmanec2004} observed photometrically 
with a main period of 4.797\,h \citep{handler2013}. 

{\changed 
We retrieved the archival \xmm\ observations of $\kappa$ Sco 
(see Table\,\ref{tab:obs}) and reprocessed them using the most 
recent data analysis software (SAS 14.0.0) and calibration  files. The source 
$\kappa$\,Sco has a visual magnitude $m_{\rm V}=2.4$, therefore the thickest
optical light  blocking filter was used in observation. For 
the \xmm\ detectors operating in full-frame mode  and with 
the thick filter, no optical loading is expected for stars 
with  $m_{\rm V}>-2$\,mag (for the PN camera). 

Unfortunately, the observations were affected 
by the high background from proton flares.  The observations were filtered
using good time intervals resulting in a useful exposure time of 
44\,ks. Here, \object{$\kappa$ Sco} is sufficiently isolated without nearby 
X-ray sources. From the region with radius $25''$ and centered on the 
coordinates of \object{$\kappa$ Sco,} we extracted the EPIC PN light 
curve in 0.2--10.0\,keV energy band and restricting the patterns 
to single and double events. Background regions were carefully 
selected on the same CCD. The net light curves with different time 
binnings  were produced using the {\it epiclccorr} task.}  
As a next step, the X-ray light curves were tested for variability.  
As an example, in the full band and binned with a 1000\,s light curve, 
the probability value
$P \approx 0.83$ is not significant even with $\alpha = 0.1$. 
Thus, the null hypothesis about the X-ray light curve from \object{$\kappa$ Sco} 
being constant cannot be rejected.  No useful information about the presence 
of periodic variability could be obtained on the basis of Fourier 
periodograms, owing to the limited quality of the data.   

\subsection{X-ray variability of \object{$\alpha$ Vir}}

\object{Spica} ($\alpha$ Vir) is a spectroscopic binary with a period of
$\sim 4$\,d in an eccentric orbit.  The primary component (B1.5IV-V)
is classified as a \bcep-type variable, but its pulsational behavior 
is not stable \citep[see discussion in][]{palate2013}. 

Spica was observed by {\em Chandra} LETG for 90180\,s in 2005-03-15
(\cxo\ observation identifier 4509).
{\changed We retrieved the X-ray light 
curves (see Fig.\,\ref{fig:lc}) from the TGCat and tested them for
variability.  As mentioned above, there is no concern about optical or
UV light leakage since we are not using the zeroth-order photons.  For
extremely high UV flux sources, the LETG coarse support structure acts as a
crude disperser, and it can sometimes  produce
spurious signal for LETG used with HRC-S \citep{Drake:2010}, but we
see no such evidence in this LETG/ACIS-S spectrum.}

Based on the $\chi^2$-test, the chance probability value $P\approx 0.19$ 
is not significant. We went on to conduct a period search, but all 
peaks in the 
Fourier periodogram are statistically insignificant (see Fig.\,3).  
No useful information on the X-ray variability was obtained from 
other statistical tests either. Moreover, in the case of $\alpha$ Vir
with its unstable pulsational behavior in order to reach
any conclusions about the correlation between stellar oscillations
and X-ray variability, it would be necessary to 
observe the star simultaneously in X-rays and in optical.

The {\em Chandra} observations of \object{$\alpha$ Vir} lasted for about 
25\,h, thus covering about a quarter of the orbital period of this binary star. 
Although no optical eclipses have 
been observed in this system \citep{palate2013}, an influence 
of binarity on the X-ray light curve cannot be excluded a priori.   
 
%=============================================================
\begin{figure}[t!]
\centering
\includegraphics[width=0.7\columnwidth]{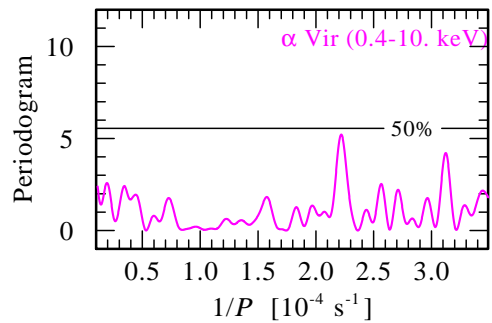}
\caption{Power density spectrum based on the Fourier transform 
\citep{scargle1982, horne1986} of X-ray light curve (0.4-10.0\,keV) 
of $\alpha$ Vir. The false alarm probability at 50\%\ level is marked.}
\label{fig:statspic}
\end{figure}
%------------------------------------------------------------- 
 
\section{Discussion}

{\changed Our study of a small 
sample of X-ray lightmcurves from \bcep-variables 
did not reveal any periodic variability. However, the X-ray 
count rate of the
only star where the X-ray pulsations are clearly detected, \xicma, 
is much greater than for the other stars considered here. It is 
important to assess what the upper limit is on an amplitude of modulation 
that could have been detected in our sample stars with the available 
data. 

We therefore wish to study a star with a periodic signal of relative amplitude $A$ 
and $N$ counts per pulsational period. The noise level can be 
approximated as $2\sqrt{N/n_{\rm p}}$, where $n_{\rm p}$ is the 
number of periods. We assume that a significant detection must be 
a factor $4\times$ higher than the noise level. Thus, we can derive 
the minimum detectable amplitude for a given observation. The results 
are shown in Table\,\ref{tab:res}. Only for \object{$\kappa$\,Sco} the 
sensitivity of observations was marginally sufficient to detect a 
variability on a level of about 10\%. The data sets for \object{$\beta$ Cru} and 
\object{$\alpha$ Vir} are not good enough to uncover periodic variability on
a level similar to that seen in \object{$\xi^1$\,CMa} even if it were present.  }

While we did not found any {\em \emph{periodic pulsations}}, our analysis showed that 
some of the sample stars show marginal evidence of X-ray variability. 
For some \bcep-type variables, such variability has also been previously 
reported in the literature.  

Significant stochastic X-ray variability is expected in the models of 
X-ray generation in stellar winds \citep{feldmeierb1997,osk2001}. 
However, the most sensitive X-ray observations of an O-type star 
{\changed (the \xmm\ observations of \object{$\zeta$\,Puppis} with the total exposure 
time of about 1\,Ms)} revealed a surprising lack of short time-scale stochastic 
variability \citep{naze2013}. 

On the other hand, the evidence is accumulating that the X-ray variability 
associated with stellar rotation may be present in massive star winds.
\citet{berg1996} observed the supergiant \object{$\zeta$ Puppis} 
for 11 days and found modulations with a frequency of 1.44\,d$^{-1}$ 
in the H$\alpha$ line profiles, 
as well as in the X-rays in the 0.9-2.0\,keV band. They reported 
an amplitude of X-ray variability of 6\%\ and suggested that 
these variations are periodic. This suggestion was not confirmed by 
later studies based on {\em ASCA} and \xmm\
observations \citep{osk-a2001, naze2013}. \citet{how2014} reported the detection 
of a period $P = 1.780$\,d  in optical light curves of \object{$\zeta$ Puppis} 
and suggested that this period may be due stellar pulsations.
While indeed the X-ray modulations on a  time  scale of a
day are present in \object{$\zeta$ Puppis}, no coherent periodicity in 
X-ray emission has been confirmed for this object so far.

Periodic modulations of X-ray emission 
were  reported for the O-type dwarf \object{$\zeta$\,Oph} \citep{osk-a2001}.
However, also in this case, we lack any information on the repeatability of X-ray 
modulations. Two other O-type stars, \object{$\xi$ Per} \citep{massa2014} and 
\object{$\zeta$ Ori} 
(A. Pollock, private communication), also show modulations of X-ray emission 
on a time scale of days. Interestingly, \xmm\ observations of a Wolf-Rayet 
type star, \object{WR\,6} showed that its X-ray light curve changes 
{\em \emph{quasi-periodically}} 
\citep{ign2013}. We speculate that, similar to WR\,6, in the OB-type stars 
the X-ray variability owing to the rotation has quasi-periodic character. 

A small fraction of B-type stars possess strong magnetic fields 
\citep[e.g.,][and references therein]{hub2006,mor2014}. 
If a magnetic field has a dipole configuration,  it can channel  
the  wind toward the magnetic equator, where wind streams from opposite  
hemispheres collide and produce a strong shock 
\citep[magnetically confined wind shock (MCWS),][]{bma1997,ud2014}. 
The MCWS predicts the X-ray variability of magnetic massive stars 
associated with the stellar rotation due to the occupation 
of X-ray emitting regions by the cold torus of matter accumulated 
in the equatorial plane  \citep{don2001, don2002}. Such
X-ray modulations on a rotational time scale are sometimes observed 
\citep[e.g.,][]{gag1997,pil2014}; however, this is not
ubiquitous behavior among  magnetic massive stars  -- some of these 
well-known objects, such as \object{$\tau$\,Sco}, do not show any evidence of X-ray 
variability over the stellar rotation period \citep{ign2010}. 

The data sets that we consider in this paper do not cover 
long enough time intervals to test the X-ray light curves for modulations 
due to stellar rotation or orbital motion. Thus, it is not clear what the 
true nature is of X-ray variability observed in some of our sample stars. 

However, in none of our sample stars, \object{$\beta$ Cru}, 
\object{$\kappa$ Sco}, or \object{$\alpha$ Vir}, 
do we find statistically significant evidence for periodic
X-ray variability. Up to now, the unambiguous presence of X-ray pulsations 
with the same period and in phase with the optical light curve has
only been detected in the magnetic star \xicma. It remains to be seen whether 
this object is exceptional or if new, better quality observations will reveal 
similar properties in other \bcep-variables.

\begin{acknowledgements}
We are grateful to the referee for useful and constrictive 
comments. This research made use of the Chandra 
Transmission Grating Catalog and archive (http://tgcat.mit.edu), 
the SIMBAD database,
operated at the CDS, Strasbourg, France, and NASA's Astrophysics 
Data System.  LMO acknowledges support from DLR grant 50 OR 1302.
\end{acknowledgements}

%\bibliographystyle{aa}
%\bibliography{magbs.bib}
%-------------------------------------------------------------------

\end{document}